\documentclass[aps,prl,twocolumn]{revtex4}

\usepackage {graphicx}
\usepackage{amsmath}
\usepackage{amssymb}

\input epsf

\usepackage{latexsym}
\usepackage{amsbsy}
\usepackage{amsfonts}
\usepackage{amssymb}

\begin{document}
\def\Z{\hbox{{\sf Z}\kern-0.4em {\sf Z}}}

\title{
Charge Density Bounds in Superconducting States  \\
of Strongly Correlated  Systems }
\author{Alexander P. Protogenov}
\affiliation{Institute of Applied Physics of the RAS, Nizhny Novgorod 603950}

\begin{abstract}
Charge density bounds of knotted and linked vortex states in 
two-component Ginzburg-Landau model are considered. 
When the mutual linking number of vector order parameter vortex lines 
is less than the Hopf invariant, these states have the lower-lying energies. 
It is shown that a set of local minima of free energy contains new classes 
of universality which exist in a hole density range limited at both finite ends. \\[5pt] PACS
numbers: 74.20.De, 74.20.Mn, 11.15.Tk, 11.10.Lm, 11.27.+d 
\end{abstract}

\maketitle

{\em Introduction.} -- 
A tangle of vortex filaments is a system which attracts  
attention due to several reasons. Along with the coherent state, which is the background 
of this vortex field distributions, the filament system also contains a disorder combination 
due to free motion of its fragments and a topological order because of 
the effects of knotting and linking of its separate parts. The degree of this order can be characterized 
by the linking number of vortices. Persistent linked field  configurations along with a 
large value of interaction energy present the basic content of what we mean when talking about 
the origin of strong correlations. 

The study of soft condensed matter, whose universal behavior is determined by 
topological characteristics, is an active area of research in 
physics \cite{M,Vol,Mm,Ms} and beyond \cite{Kbmsds}. 
Indeed, the dynamics of entangled vortex states is recognized as one of the most challenging 
problems of modern condensed matter physics. 
The emergence of novel analytical methods \cite{Vol,Mm,Ms} 
to treat nonlinear field equations create now an unique opportunity to understand the 
dynamics of entangled vortex states.     

The vortex dynamics can be understood in detail in the framework of basic models, 
such as the  Ginzburg-Landau functional or more complicated multi-component systems.  
Main theoretical approaches for analyzing universal behavior in these models are based 
on the methods of the topological field theory \cite{W} and nonlinear dynamics \cite{IALK}. 
The universality classes are defined both by symmetry and topological characteristics 
of the background nonlinear fields, while a system with infinite number of 
degrees of freedom contains in the ground state only the finite number 
of symmetrically invariant physical states.

The aim of this paper is to elucidate the physical mechanism leading to formation 
of the charge density range bounds for superconducting states with a set of numbers 
determining the knotting and linking degree of the fields that take part in the description 
of the coherent state. 
The use of the Ginzburg-Landau functional and the methods of the topological gauge 
field theory gives us the hope that the obtained answers are universal. 
A combination of novel 
methods will 
make it possible to advance in 
description of the processes which are determined by the 
contribution of topological excitations.

{\em Model.} -- We will use the Ginzburg-Landau free energy 
\begin{multline} 
F =\int d^{3}x \, \biggl[\sum_{\alpha} \frac{1}{2m} \left| \left(\hbar \partial_{k} + 
i \frac{2e}{c} A_{k}\right) \Psi_{\alpha} \right|^2   \\  
+ \sum_{\alpha}
\left(-b_{\alpha}|\Psi_{\alpha }|^{2} + 
\frac{c_{\alpha }}{2}|\Psi_{\alpha }|^4 \right) + \frac{\bf B^{2}}{8\pi} \biggr]
\end{multline}
with a two-component order parameter
\begin{equation}
\Psi_\alpha = \sqrt{2m} \,\rho \,\chi_\alpha , \, \, \, \,   
\chi_\alpha=|\chi_\alpha|e^{i \varphi_\alpha} \, , 
\end{equation}                                                                            
satisfying the $CP^1$ condition, $|\chi_{1}|^{2} + |\chi_{2}|^{2}=1$. 
This model is used in the context of the two-gap superconductivity \cite{Bfn} and in the 
non-Abelian field theory \cite{Fn1,Cho}. 

It has been shown in paper \cite{Bfn} that there exists an exact mapping of the model (1),
(2) into the following version of ${\bf n}$-field model:
\begin{multline}
F =\int d^{3}x \biggl[\frac{1}{4}\rho^{2}\left(\partial_{k}{\bf n}\right)^{2} + 
\left(\partial_{k}\rho \right)^{2} + \frac{1}{16}\rho^{2}{\bf c}^{2} \\ 
+ \left(F_{ik} - H_{ik}\right)^{2} + V(\rho, n_{3})\biggr] \, .
\end{multline}
To write down Eq. (3), dimensionless units and gauge invariant order parameter fields 
of the unit vector ${\bf n}={\bar \chi}{\boldsymbol{\sigma}}\chi$, 
and the velocity  ${\bf c}={\bf J}/\rho^{2}$ have been used. Here 
$\bar \chi = (\chi_{1}^{\ast}, \chi_{2}^{\ast})\,$ and $\boldsymbol{\sigma}$ - Pauli matrices. 
The effective coupling constant $512e^{2}/\hbar c$ in this case \cite{Bfn} has the order of unity. 
The full current ${\bf J} = 2\rho^{2}({\bf j} - 4{\bf A})$ has a paramagnetic 
$\left({\bf j}=i[\chi_{1}\nabla \chi_{1}^{\ast} - c.c. + (1 \to 2)]\right)$ and 
a diamagnetic $(-4{\bf A})$ parts. Besides, in Eq. (3) 
$F_{ik} = \partial_{i}c_{k} - \partial_{k}c_{i}$, $H_{ik} = {\bf n}\cdot[\partial_{i}{\bf n}\times \partial_{k}{\bf n}]:= \partial_{i}a_{k} - \partial_{k}a_{i} $. 

Setting in Eq. (3) ${\bf c} = 0$ we come back to the model of \cite{Fn2}. The numerical study of
the knotted configurations of ${\bf n}$-field in this model has been done 
in \cite{Gh,Bs,Hs}. The lower energy bound in this case
\begin{equation}
F \geqslant 32\pi^{2}|Q|^{3/4}
\end{equation}
is determined \cite{Vk,Kr,Ward} by the Hopf invariant, 
\begin{equation}
Q = \frac{1}{16\pi^{2}}\int d^{3}x \, \varepsilon_{ikl}a_{i}\partial_{k}a_{l} \, .
\end{equation}

At compactification $R^{3} \to S^{3}$ and ${\bf n} \in S^{2}$, 
the integer $Q \in \pi_{3}(S^{2}) = \Z$ shows the degree of linking or knotting
of filamental manifolds ${\cal M}\/ \in S^{3}$, where the vector field ${\bf n}(x,y,z)$ is defined. 
In particular, for two linked rings (Hopf linking) Q=1, for the trefoil knot Q=6 and etc. 
Significant point \cite{Aw} is as follows:  
$\pi_{3}(CP^{M}) = 0$ at $M > 1$ and $\pi_{3}(CP^{1}) = \pi_{3}(S^{2}) = \Z$.  
In the latter case the order parameter (2) is two-component one \cite{Bfn} and linked or knotted 
soliton configurations are labeled by the Hopf invariant (5).

{\em Energy bounds.} -- Let us assume that $\rho = \rho_{0}$ can be find from the 
minimal value $V(\rho_{0})$ of the potential $V(\rho)$, 
but the velocity ${\bf c}$ does not equal zero. Equation (3)
in this case has the following form:
\begin{multline}
F = F_{n} + F_{c} - F_{int} = \int d^{3}x \biggl[\left(\left(\partial_{k}{\bf n}\right)^{2} + H_{ik}^{2}\right) \\ 
+ \left(\frac{1}{4}{\bf c}^{2} + F_{ik}^{2}\right) - 
2F_{ik}H_{ik}\biggr] \, .
\end{multline}

It is seen from Eq. (6) that a superconducting state with $|{\bf c}| \ll 1$ has the energy
which is less than the minimum in Eq. (4). 
To find the lower free energy bound in the superconducting state with ${\bf c} \neq 0$,  
it has been proved \cite{Pv} that the following inequality for free energy parts takes place 
\begin{equation}
F_{n}^{5/6}\,F_{c}^{1/2} \geqslant (32\pi^{2})^{4/3}|L| \, ,    
\end{equation} 
where
\begin{equation}
L = \frac{1}{16\pi^{2}}\int d^{3}x \, \varepsilon_{ikl}c_{i}\partial_{k}a_{l} 
\end{equation} 
is the degree of mutual linking \cite{Zk} of the velocity ${\bf c}$ lines and of the 
magnetic field ${\bf H} = [\nabla \times {\bf a}]$ lines. 
Quantities like (5), (8) arise as first integrals in the theory of an ideal or barotropic 
fluid \cite{Zk}. In general, the integrals in Eqs. (5), (8) could be defined by the asymptotic 
linking numbers \cite{Ak}.  

It follows from the Schwartz-Cauchy-Bunyakovsky inequality that 
$2F_{n}^{1/2}\cdot F_{c}^{1/2} \geqslant F_{int}$.  
Setting the boundary value $F_{int}$ into Eq.(6), we get $F_{min}=(F_{n}^{1/2} - F_{c}^{1/2})^{2}$. 
The mimimum value $32\pi^{2}Q^{3/4}$ of the 
function $F_{n}$ and Eq. (7) lead \cite{Pv} to  
\begin{equation}
F  \geqslant 32\pi^{2}\,|Q|^{3/4}\,(1 - |L|/|Q|)^{2} \, \, .  
\end{equation}
The trivial case $Q=0$ should be considered after the limit $L=0$. 
Let us pay also an attention to the 
self-dual relation $F_{n} = F_{c}$ which follows from $F_{min}$. 

In general, the Hopf-Chern-Simons matrix,  
\begin{equation}
K_{\alpha \beta} = 
\frac{1}{16\pi^{2}}\int d^{3}x \, \varepsilon_{ikl}a_{i}^{\alpha}\partial_{k}a_{l}^{\beta} = 
\left( \begin{array}{cc} 
Q & L^{\prime}  \\ 
L & Q^{\prime}    
\end{array} \right)  \, .
\end{equation}
can be defined in our case by linking numbers $Q \in \Z$ of spin and spin-charge 
($\{L, Q^{\prime}\} \notin \Z $ at all) degrees of freedom.  
In Eq. (10), $K_{\alpha \beta}$ is a symmetrical matrix $(L^{\prime} = L)$ \cite{Ak} and 
$a_{i}^{1} \equiv a_{i}, \, a_{i}^{2} \equiv c_{i}$. 
Following the approach of \cite{Pv} which is based on employ 
of the H\"older inequality chain  as well as on the Ladyzhenskaya \cite{Lad}  
inequality, one can find that 
\begin{equation}
F_{n}^{1/2}\,F_{c}^{5/6} \geqslant (16\pi^{2})^{4/3}|L^{\prime}| \, .    
\end{equation}  
The distiction of the coefficient in Eq. (11) from Eq. (7) results from the charge $2e$ of pairs (due to 
the coefficient $1/4$ in the firt term of the free energy part $F_{c}$).  

It follows from Eq. (9) that for all numbers $L < Q$ the energy 
of the ground state is less than that in the model described in \cite{Fn2}, for which the 
inequality (4) is valid. 
The origin of the energy decrease can be easily understood. Even under the conditions
of the existence of the paramagnetic part ${\bf j}$ of the current ${\bf J}$, 
the diamagnetic interaction in the superconducting state consumes its own energy of the 
current and a part of the energy relating to the ${\bf n}$-field dynamics for all state 
classes with $L \leqslant Q$. 

{\em Charge density}. -- In the context of the high-temperature superconductivity 
problems \cite{SBL,Lnnw,Lw,Clmn}, Eq. (2) has the sense of the condition of the correlated or 
linked factorization of charge and spin degrees of freedom (from the gauge invariant order parameter 
point of view (3)). In fact, the constant value of the charge density 
plays the role of a tuning parameter of the system. In this case, the order parameter ${\bf n}$ describes the distribution of spin degrees of freedom, while the order parameter 
${\bf c}$ contains the contribution both of spin degrees 
of freedom $\chi_{\alpha}$ in the paramagnetic part of the current and of the 
U(1) charge degrees of freedom in the diamagnetic part of the current ${\bf J}$. 
Due to such mixing, the gauge invariant order parameters ${\bf n}$
and ${\bf c}$ differ principally and topologically. 
If the mutual linking index is defined by the 
current ${\bf J}$, the measure in Eq. (8) depends on the charge density.
The vector ${\bf c}={\bf J}/\rho^{2}$ is normalized to the charge density and, as 
distinct from ${\bf n}$, belongs to the noncompact manifold. 
This leads to the fact that the Hopf numbers for it 
are not integer valued ($(L, Q^{\prime}) \notin \Z$), and the linking numbers 
are included into the factor of Eq. (9) in the form of the ratio $L/Q$.  

Let the parameter $\rho_{0}$ change in some range. Since all terms in Eq. (3)  
are of the same order, the velocity ${\bf c}$ and therefore the number $L$ decrease     
as $\rho_{0}$ increases. In this case, if $L$ is rather small, the smallest 
superconducting gap goes down with the growth of $Q$ at the background of large value 
$32\pi^{2}Q^{3/4}$ of the spin gap. 

As $\rho_{0}$ decreases, the following effect takes place. The radius $R$ of the 
compactification $R^{3} \to S^{3}$ being proportional to $\rho^{-1}_{0}$, 
increases till it exceeds a certain critical value, $R_{cr}$. 
At $R > R_{cr}$ the Hopf map \cite{KM} is unstable relative to infinitesimal perturbation of 
vortex linked field distribuions. As a result, the $U(2)$ symmetry associated 
with identical Hopf map is spontaneously broken. This means that the topological 
solitons, instead of being spread out over the whole of $S^{3}$, are then localized around
a particular point (the base point of the stereographic projection) and collapse
into localized structures \cite{Ward}. This picture corresponds to the existence of an optimal value  
of the relation $L/Q$. Under such a key condition of the restoration of compactness  of the base 
manifold, inhomogeneous superconducting state appears. As the parameter $\rho_{0}$  
decreases further, the free energy $F$ due to the inhomogeneity term $(\partial_{k} \rho)^{2}$ 
increases again and suppresses the superconducting gap. 

Up to now the vector ${\bf A}$ characterized the internal charge $U(1)$ gauge symmetry. 
If we apply an external electromagnetic field the vector potential ${\bf A}$ 
equals the sum of the internal and the external gauge potentials. 
As a result, due to the diamagnetism of the superconducting state, the velocity ${\bf c}$ 
decreases. Like in the case of increasing $\rho_{0}$, this leads to 
suppression of the superconducting gap. The answer to the question on the existence of full or partial 
Meisner screening in these states depends on the result of the competition of contributions 
from neutral ${\bf j}$ and charged $-4{\bf A}$ parts to the full current ${\bf J}$. 

The soft case $\rho \neq const$ both for ${\bf c}=0$ \cite{Lnnt} and ${\bf c} \neq 0$ 
arouses certain interest. It is more complicated due to some reasons and 
will be considered in a separate paper. 
The point $\sqrt{F_{c}/F_{n}} = 1/4$ where the fields of the inequalities (7), (11) 
intersect, deserve an additional careful study as well.  
The equality in Eq. (9) under this remarks 
should be understood as an ideal limit depending on topologial characteristics of 
knots and links only.  

\begin{figure}
\includegraphics[width=78mm, height=43mm ]{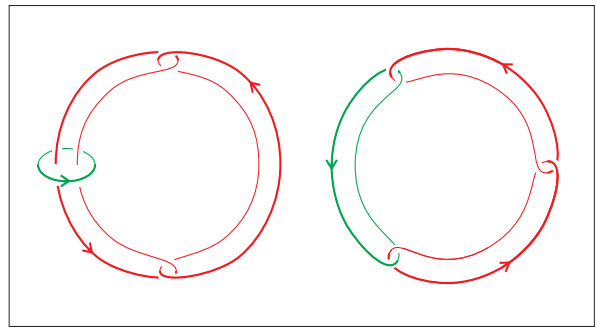}  
\caption{ 
Examples of ${\bf n}$- and ${\bf c}$-field line configurations in the form of Borromean rings 
with vanishing pairwise linking numbers $K_{\alpha \beta}$ (on the l.h.s.)  
and in the form of the closed chain of correlated charge-spin semion loops with $Q=1$ (on the r.h.s.). 
}
\label{bor}
\end{figure}

{\em Discussion}. -- In the $(3+0)D$ case of the free energy (3), 
Hopf invariant (5) is analogous to the Chern-Simons action 
$(k/4\pi)\int dt\,d^{2}x\,\varepsilon_{\mu \nu \lambda}
a_{\mu}\partial_{\nu}a_{\lambda}$ determining 
strong correlations of $(2+1)D$ modes \cite{Apv,Pro}  
at $k \simeq 2 $. In planar systems, this coefficient has the sense of braiding of the  
excitation world lines. In particular, for the semion $k=2$.  
Keeping in mind the relation of spatial dimensionality of 
the systems in their quantum and statistical  
descriptions, we note that the $(2+1)D$ dynamical case $k=2$ of 
the open world line ends of excitations is equivalent to the 
compact $(3+0)D$ statistical example of the Hopf linking 
$Q=1$ in Eq. (3) (see Fig. \ref{bor}).  
 
Similarly to the description of the states in the fractional quantum Hall effect 
with the filling factor $\nu = p/q \, \,$ and $p,q \in \Z $, the Eq. (9) depends on $L/Q$.  
The Hopf invariant $Q \in \Z $ enumerating gauge vakua \cite{BW}, is equivalent to the degree $q$ 
of the ground state degeneracy and $L$ plays the role 
of the filling degree $p$ of the incompressible charged fluid state. 
We have stressed already that the number $L$ is, generally, 
not integer valued. From this point of view the multiplier $(1 - L/Q)$ in Eq. (9) 
is equivalent to the Hall filling factor $1-\nu$ for holes. 
(Let us remark that the mentioned parameter $\rho_{0}^{2}$ has the sense of the particle 
density). The distinction of our system from 
the fractional quantum Hall effect state is that 
due to compressibility of the superconducting state, where the charge gauge $U(1)$ 
symmetry is broken, the effective number $L$ of the charge degrees of freedom 
is continuous. It plays the role of an interpolation parameter \cite{Gr} which connects 
sectors with $L \notin \Z $ and $L \in \Z $  
in one and the same universality class of Pfaffian states.   
Apparently it is true for all Pfaffian-like states with $Q >1$ and numbers $L < Q$. 

In conclusion, we have studied the charge density bounds of the superconducting 
states using the $CP^{1}$ version of Ginzburg-Landau model under the conditions of the
existence of linking and knotting phenomena of the ${\bf n}$- and ${\bf c}$-fields 
being the gauge invariant order parameters of the considered system. 
We have shown that the superconducting states exist in the finite range 
$\rho_{1}^{2} < \rho_{0}^{2} < \rho_{2}^{2}$ of the charge density values. 
In this range, the field configurations with field lines characterized by semion 
values of the linking numbers are preferable.  

I would like to thank A.G. Abanov, E.A. Kuznetsov and G.E. Volovik 
for advices, V.F. Gantmakher for the crucial remark, and L.D. Faddeev, G.M. Fraiman, A.G.
Litvak, V.A. Verbus for useful discussions. This work was supported in part
by the RFBR under the grant No. 01-02-17225.


\begin{thebibliography}{99}

\bibitem{M} H.K. Moffatt, Nature {\bf 347}, 367 (1990).  

\bibitem{Vol} V.M.H. Ruutu, U. Parts, J.H. Koivuniemi, M. Krusius, E.V. Thuneberg, and   
G.E. Volovik, JETP Lett. {\bf 60}, 671 (1994).  

\bibitem{Mm} Yu.G. Makhlin, T.Sh. Misirpashaev, JETP Lett. {\bf 61}, 49 (1995). 

\bibitem{Ms} M.I. Monastyrsky, P.V. Sasorov, Sov. Phys. JETP {\bf 66}, 683 (1987).  

\bibitem{Kbmsds} V. Katritch, J. Bednar, D. Michoud, R.G. Scharein, J. Dubochet, and A. Stasiak, 
Nature {\bf 384} 142 (1996).

\bibitem{W} E. Witten, Commun. Math. Phys. {\bf 121}, 351 (1989). 

\bibitem{IALK} I. Aranson, L. Kramer, Rev. Mod. Phys. {\bf 74}, 99 (2002).  

\bibitem{Bfn} E. Babaev, L.D. Faddeev, A.J. Niemi, Phys. Rev. B {\bf 65}, 100512 (2002); 
cond-mat/0009438.   

\bibitem{Fn1}  L.D. Faddeev,  A.J. Niemi,  Phys. Lett. {\bf B525}, 195 (2002).

\bibitem{Cho} Y.M. Cho,  Phys. Rev. Lett.  {\bf 87}, 252001 (2001); hep-th/0110076.  

\bibitem{Fn2}  L.D. Faddeev,  A.J. Niemi,  Nature {\bf 387}, 58 (1997). 

\bibitem{Gh} J. Gladikowski, M. Hellmund, Phys. Rev. {\bf D} {\bf 56}, 5194 (1997). 

\bibitem{Bs} R.A. Battye, P.M. Sutcliffe, Phys. Rev. Lett. {\bf 81}, 4798 (1998). 

\bibitem{Hs} J. Hietarinta, P. Salo, Phys. Lett. {\bf B451}, 60 (1999). 

\bibitem{Vk}  A.F. Vakulenko, L.V. Kapitansky, Sov. Phys. Dokl. {\bf 24}, 433 (1979). 

\bibitem{Kr} A. Kundu, Yu.P. Rubakov, J. Phys. A {\bf 15}, 269 (1982).   

\bibitem{Ward} R.S. Ward, Nonlinearity {\bf 12}, 1 (1999), hep-th/9811176. 

\bibitem{Aw} A.G. Abanov, P.W. Wiegmann. Geometrical phases and quantum numbers of 
solitons in nonlinear sigma-models, hep-th/0105213.

\bibitem{Pv} A.P. Protogenov, V.A. Verbus, JETP Lett. (to be published). 

\bibitem{Lad} O.A. Ladyzhenskaya. The mathematical theory of viscous 
incompressible flow, Gordon and Breach, 1969. 

\bibitem{Zk} V.E. Zakharov, E.A. Kuznetsov, Phys. Usp. {\bf 40}, 1087 (1997). 

\bibitem{Ak} V.I. Arnold, B.A. Khesin. Topological Methods in Hydrodynamics. Appl. Math. Sci. 
{\bf 125}, Chapt.3. 

\bibitem{SBL} M. Sigrist, D.B. Bailey, and R.B. Laughlin, Phys. Rev. Lett. {\bf 74} 3249 (1995). 

\bibitem{Lnnw} P.A. Lee, N. Nagaosa, T.K. Ng, and X.G. Wen, Phys. Rev. {\bf B 57}, 6003 (1998).  

\bibitem{Lw} P.A. Lee and X.G. Wen. Vortex structure in underdoped cuprates, cond-mat/0008419. 

\bibitem{Clmn} S. Chakravarty, R.B. Laughlin, D. Morr, C. Nayak. Hidden order in the cuprates, cond-mat/0005443.  

\bibitem{KM} E.A. Kuznetsov, A.V. Mikhailov, Phys. Lett. A {\bf 77} 37 (1980).  

\bibitem{BW} P. van Baal, A. Wipf. Classical gauge vacua as knots, hep-th/0105141. 

\bibitem{Lnnt} M. L\"ubcke, S.M. Nasir, A. Niemi, and K. Torokoff, Phys. Lett. 
{\bf B 534} 195 (2002); hep-th/0106102.  

\bibitem{Apv} L.A. Abramyan, A.P. Protogenov, V.A. Verbus, JETP Lett. {\bf 69}, 887 (1999).  

\bibitem{Pro} A.P. Protogenov, JETP Lett. {\bf 73}, 255 (2001).

\bibitem{Gr} N. Read, and Dmitry Green, Phys. Rev. B {\bf 61}, 10267 (2000). 
 
\end{thebibliography}
\end{document}